\documentclass[12pt]{article}

\usepackage{scicite}
\usepackage{graphicx}
\usepackage{braket}
\usepackage{hyperref}

\usepackage{times}

\newenvironment{sciabstract}{%
\begin{quote} \bf}
{\end{quote}}

\newcounter{lastnote}

\title{Entropy and complexity unveil the landscape of memes evolution}

\author
{
Carlo M. Valensise,$^{1\ast}$ Alessandra Serra,$^{2}$ Alessandro Galeazzi,$^{3}$  \\
Gabriele Etta,$^{4}$  Matteo Cinelli,$^{5,6}$  Walter Quattrociocchi$^{4}$\\
\\
\normalsize{$^{1}$Enrico Fermi Research Center, Piazza del Viminale, 1 — 00184 Roma (IT),}\\
\normalsize{$^{2}$Tuscia University - DISTU}\\
\normalsize{Department of Modern Languages and Literatures, History, Philosophy and Law Studies}\\
\normalsize{Via S. Carlo, 32 — 01100 Viterbo (IT),}\\ 
\normalsize{$^{3}$University of Brescia, Via Branze, 59 — 25123 Brescia (IT)}\\
\normalsize{$^{4}$Sapienza University of Rome — Department of Computer Science,}\\
\normalsize{ Viale Regina Elena, 295 — 00161 Roma (IT)}\\
\normalsize{$^{5}$Italian National Research Council — Institute for Complex Systems}\\
\normalsize{Via dei Taurini 19, 00185 Roma (IT)}\\
\normalsize{$^{6}$Ca’ Foscari, University of Venice} \\
\normalsize{Department of Environmental Sciences, Informatics and Statistics}\\
\normalsize{Via Torino 155, 30172 Mestre (IT)}\\

\normalsize{$^\ast$To whom correspondence should be addressed; E-mail:  carlo.valensise@cref.it}

}

\date{}

\begin{document} 


\baselineskip24pt


\maketitle


\begin{sciabstract}
On the Internet, information circulates fast and widely, and the form of content adapts to comply with users' cognitive abilities.
Memes are an emerging aspect of the internet system of signification, and their visual schemes evolve by adapting to a heterogeneous context. 
A fundamental question is whether they present culturally and temporally transcendent characteristics in their organizing principles. 
In this work, we study the evolution of 2 million visual memes from Reddit over ten years, from 2011 to 2020, in terms of their statistical complexity and entropy.  
We find support for the hypothesis that memes are part of an emerging form of internet metalanguage: on one side, we observe an exponential growth with a doubling time of approximately 6 months; on the other side, the complexity of memes contents increases, allowing and adapting to represent social trends and attitudes.
\end{sciabstract}


\section*{Introduction}
Social media radically changed the way we consume information and interact online \cite{del2016spreading,schmidt2018polarization,sunstein2017republic}.
Online interactions, indeed, influence social dynamics by favoring the formation of homophilic groups around shared narratives and attitudes and thus bursting group polarization \cite{del2016echo,cinelli2021echo,bail2018exposure}.
In this scenario, multimedia content such as videos, photos, and pictures represents an essential portion of online communication, especially within social media platforms.
Online communication can be read through the lenses of Dawkins' \textit{cultural memes} \cite{Dawkins2016}, whose definition applies to almost all online information vehicles. Cultural memes represent a unit of cultural information transmitted and replicated; writing posts, sharing personal videos, expressing "likes" are examples of this concept.
While Dawkins' model of cultural evolution is nowadays considered insufficient to comprehend the complex cultural phenomena of information transmission \cite{Deacon_1999,Sebeok_2000,Cannizzaro_2016,Fomin_2019}, its evolutionary pattern still represents a valid basis for describing fundamental features of memes diffusion.
In this work, we investigate the role and evolution of a particular kind of cultural meme, namely template images that undergo modifications or get some text overlapped, conventionally referred to just as \textit{memes}. In the following, we adopt this convention.
According to Dawkins' hypothesis, cultural memes \cite{Distin2004} are characterized by the three essential elements of evolutionary theory: replication, variation, and selection. 
In the case of visual memes, the replication mechanism is self-evident. It consists of modifying an image, e.g., with some text, to represent a given situation. Moreover, replication of memes is facilitated by their consistency with other cultural memes present in the online environment, such as short videos, pictures, or short texts.
Variation is an intrinsic feature of visual memes. Indeed, new memes are continuously created to target funny situations or jokes about political or societal events and compete for users' attention flowing across online communities.
Finally, selection occurs when a meme cannot attract human attention nor adapt to transmit new contents and disappears adapting to the fast online environment. 
Among the online cultural memes that underwent relatively strong selection, we find, for example, blogs and discussion forums that have been replaced mainly by online social media; similarly, the emoji's introduction strongly reduced the use of \textit{ascii art} symbols.

So far, a large body of research quantitatively investigated the features of different online cultural memes, not limited to images. Textual memes were analyzed by Leskovec et al. \cite{Leskovec2009} as a proxy for the cycle of online news consumption. Ienco et al. \cite{ienco2010meme} studied the problem of ranking memes, i.e., selecting those memes to be displayed to users to maximize the network activity on the platform. Romero et al. \cite{Romero2011} studied online memes propagation in the form of Twitter hashtags. Bauckhage \cite{bauckhage2011insights} investigated the epidemic dynamics of 150 famous memes, applying models from mathematical epidemiology to account for the growth and decline of visual memes. Ratkiewicz and coworkers \cite{Ratkiewicz_2011} developed a framework for analyzing the diffusion of politics-related tweets. Weng et al. \cite{Weng2012} studied meme virality through an agent-based approach, accounting for the limited attention each user can spend in online environments. In \cite{weng2014predicting} the popularity of memes is correlated with the underlying network community structure. In \cite{Ferrara2013} clustering techniques are applied to identify text-based memes, leveraging the content, metadata, and network structure of social data. Coscia \cite{Coscia2014} studied the popularity of memes leveraging measures of similarity between memes. In \cite{Dang2015} tri-grams are used to cluster posts from Reddit. In \cite{tsur2015don} textual memes popularity is investigated looking at linguistic features. Adamic et al. \cite{Adamic2016} explored a large corpus of textual data from Facebook modeling the propagation of information as a Yule process.
Dubey et al. \cite{dubey2018memesequencer} employed a Deep Learning architecture to process memes, extract the underlying template and explore its variations. 
An extensive analysis of visual memes is performed by Zanettou and coworkers \cite{zannettou2018origins} exploiting perceptual hashing to cluster visual memes together and explore the connections between the meme content and the communities in which it circulates. 
In \cite{Beskow2020} a deep-learning classifier for memes is proposed to explore the role of memes instead of non-meme images during elections.
These investigations developed relevant insights and tools for handling and researching the world of internet memes. Nevertheless, little attention is given to the fundamental aspects of the evolution of memes in terms of visual features and conveyed information. Eventually, no evidence is reported for the hypothesis of internet memes constituting a metalanguage of the internet \cite{Cannizzaro_2016,Fomin_2019}.

In this work, we investigate the general evolution pattern of memes as an online communication artifact. To this aim, we leverage the evolutionist approach to define and measure the evolution rate, i.e., the number of new templates that appear online per time unit, the variation rate, i.e., the number of new instances of the same template that are produced in time; this quantity is particularly relevant concerning memes' popularity. Finally, from a general perspective, we compute the trajectory of memes in the entropy-complexity plane. These measures are grounded in the physics of complex systems and have been employed to investigate painting arts \cite{Sigaki2018}, revealing a temporal pattern towards higher complexity.

The basis of this investigation is a massive dataset of 2 million Reddit memes over ten years. Each image has been classified and ascribed to a template through a Machine Learning pipeline composed by an unsupervised Deep-learning based classifier followed by a density-based clustering algorithm (see Methods). 
Our investigation shows that the memes ecosystem size is exponentially increasing, with a doubling time of approximately six months, indicating that replication is currently the leading process.
Concerning selection, we observe that memes' persistence is dominated by rapid early adoption.
The variation pattern is captured by the trajectory in the entropy--complexity plane. Similarly to what happens in painting arts, we observe a tendency towards structures with increasing visual complexity; early memes were made up of simple foreground images (e.g., animals or explicit human expressions) on plain backgrounds, while later ones involve more articulated scenes (e.g., modified movie frames).

As cultural signs, memes are strictly connected to the broader cultural system in which they are embedded. While their ultimate theoretical definition is still elusive and debated in terms of methodological frames, our results indicate that memes appear as one of the most productive and adaptable areas of digital communication, functioning as a metalanguage of cultural dynamics and evolving in progressive forms of textual complexity.

\section*{Methods}
\label{sec:methods}
\subsection*{Data Breakdown}
Reddit \cite{reddit} is an online social media platform that aggregates users in communities of interest. In the last years, it has been widely employed to perform academic research on online communities, and the number of active users on this platform is constantly increasing \cite{Medvedev_2019}.  

\begin{figure}
    \centering
    \includegraphics{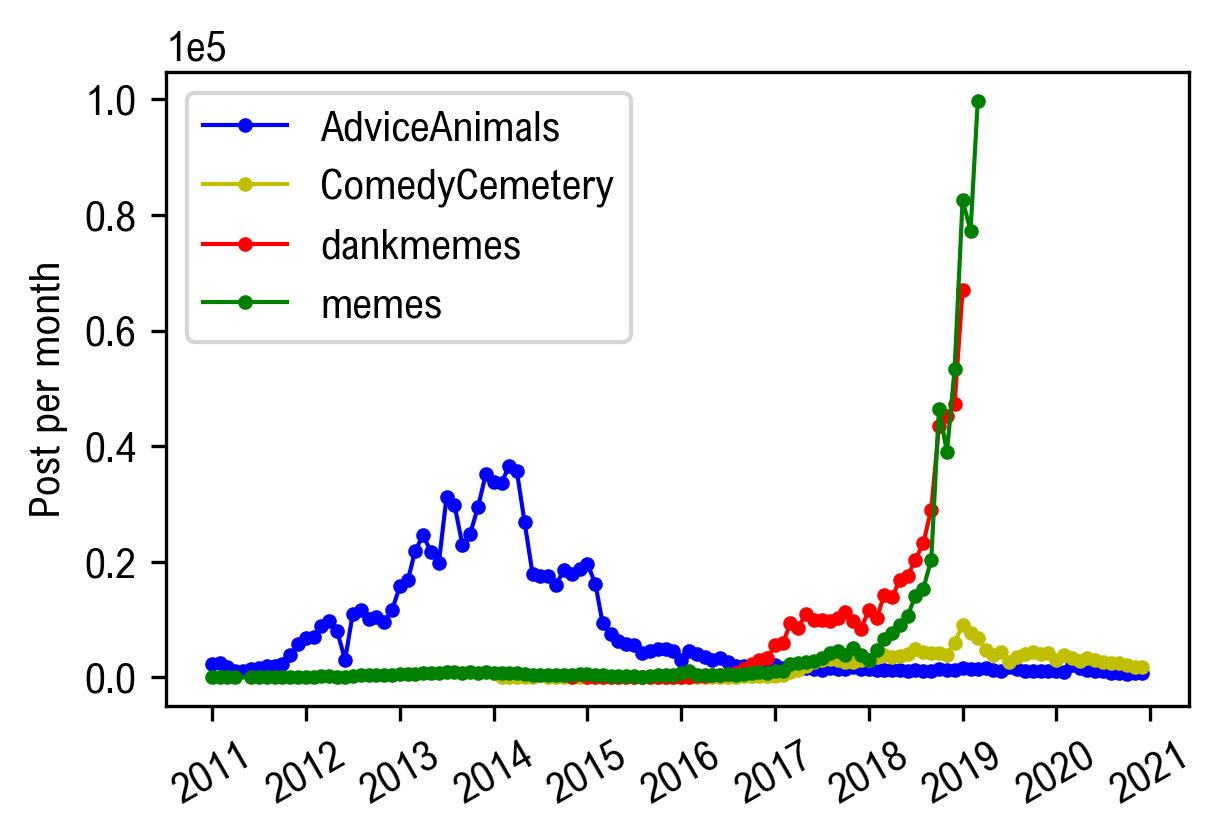}
    \caption{Dataset used in this work. The total amount of downloaded memes is about 2 millions.}
    \label{fig:dataset}
\end{figure}

The visual memes used as dataset for this study were downloaded through the Pushshift Reddit Dataset \cite{baumgartner2020pushshift}, selecting four communities (subreddits) explicitly devoted to share and discuss about memes, namely: \textit{r/AdviceAnimals}, \textit{r/memes}, \textit{r/CemeteryComedy} and \textit{r/dankmemes}. Data were collected considering a ten years window, from 2011 to 2020. In Figure~\ref{fig:dataset} the number of downloaded posts per each community is reported, as function of time. Not all the communities started their activity simultaneously. 

\subsection*{Clustering}
One of the main features of visual memes is their recurrent nature: starting from an initial template, memes are produced through text or image modifications resulting each time in a new instance that stems from the original template. 
To measure the evolution and variation rate of visual memes, it is crucial to cluster them according to the underlying template. As the collected number of memes is about two million, such a large amount of images calls for automatic classification methods. 

Our unsupervised clustering procedure is divided into two steps: first, we apply, to our knowledge, the state of the art Deep Learning implementation for unsupervised image clustering, that is called SCAN \cite{vangansbeke2020scan} (Semantic Clustering by Adopting Nearest neighbors), followed by a further clustering procedure through the HDBSCAN (Hierarchical Density-Based Spatial Clustering of Applications with Noise) algorithm \cite{Campello2013}. 

Specifically, the SCAN algorithm works in two steps. In the first one, self-supervised learning is used to train a neural network with parameters $\vartheta$ that maps images ($x_i, i=1\dots N$) into feature representations $\phi_\vartheta(x_i)\in {\rm I\!R}^d $.
Therefore, each image is represented by a vector of dimension $d$, with $d$ being the dimension of the embedding space carrying semantically meaningful information about its content. The parameters $\vartheta$ are determined by minimizing the loss function given by the distance $\delta$ between the representation of the image and the representation of their augmentations:
\[
\min_\vartheta \delta(\phi_\vartheta(x_i),\phi_\vartheta(T[x_i]))\,.
\]
The augmentation of an image may be a rotation, an affine or perspective transformation, etc.

In the second step, a neural network with parameters $\eta$ is used to classify an image $x_i$ and its nearest neighbors sampled exploiting its corresponding representation $\phi_\vartheta(x_i)$. As the second task is a classification, the output is a probability distribution over the considered classes for the given image: $\phi_\eta(x)\in [0,1]^C$, where $C$ is the number of clusters considered.

The loss function, in this case, is made of two terms
\[
\Lambda = -\frac{1}{|\mathcal D|} \sum_{x\in \mathcal D}\sum_{k\in \mathcal N_x}\braket{\phi_\eta(x),\phi_\eta(k)} + \lambda \sum_{x\in \mathcal C} \phi'^c_\eta \log(\phi'^{c}_\eta)\,, 
\]
with
\[
\qquad \phi'^{c}_\eta = -\frac{1}{|\mathcal D|} \sum_{x\in \mathcal D} \phi_\eta^c(x)\,,
\]
where $\mathcal D$ is the dataset comprising all images, $\mathcal N_x$ is the set of nearest neighbors of image $x$, $\mathcal C$ is the set of clusters, and  $\braket{\cdot}$ is the dot product. The first term aims to maximize the probability that the image and its nearest neighbors are classified in the same class. The second term avoids the formation of a single cluster containing all the images and forces the spread of the predictions uniformly across the clusters. Eventually, for each processed image, we get a representation vector of size $d=2048$ and a set to which it belongs. 

The SCAN algorithm provides us with an informative and compact representation for each image in our dataset together with a first, high-level clustering. SCAN divides the corpus into four clusters that group the visual memes into three broad and general categories: animals (two sets), humans, and others. The two clusters with animal images have been merged together. In other words, memes containing humans represent the 50\% of the corpus, followed by 25\% of animals and 25\% of other kind of contents.

To obtain a template-based clustering, we exploited HDBSCAN \cite{Campello2013} implemented by \cite{McInnes2017}. For each high-level cluster obtained from SCAN, the entire corpus of memes can be represented by a matrix whose rows are the representations $ \phi_\vartheta(x_i)$. To make the problem computationally more tractable, Principal Component Analysis was used to reduce the features' dimension from 2048 to 20. This allowed us to better fit our computational resources and did not cause any reduction in the quality of the clustering. By applying HDBSCAN to such a matrix, we were able to get a label for each image of our corpus and to separate the memes by their template. Notably, HDBSCAN can separate clusters from noisy points. Part of the corpus does not belong to any template, and therefore is marked as noise and grouped in a large "cluster of noise". Despite this suitable property of the algorithm, some clusters may result made up of images whose template is not the same for all. A purity measurement is therefore required to exclude from the analysis too heterogeneous clusters, i.e. clusters below a given purity threshold. 

In a completely unsupervised framework, the quality of clustering is in general not easy to evaluate \cite{Mehta_2019}. A quantity that is usually employed as an objective function for clustering is the average-pairwise distance $\overline{S}_k$, which evaluates the intra-cluster homogeneity \cite{Rokach}, i.e. how much each element of the cluster is, on average, similar to all the others. Its definition is given by
\[
\overline{S}_k = \frac 1{N_k^2}\sum_{x_i,x_j \in \mathcal C_k}^{N_k} || \phi_\vartheta(x_i)-\phi_\vartheta(x_j) ||_2^2\,.
\]

\begin{figure}
    \centering
    \includegraphics{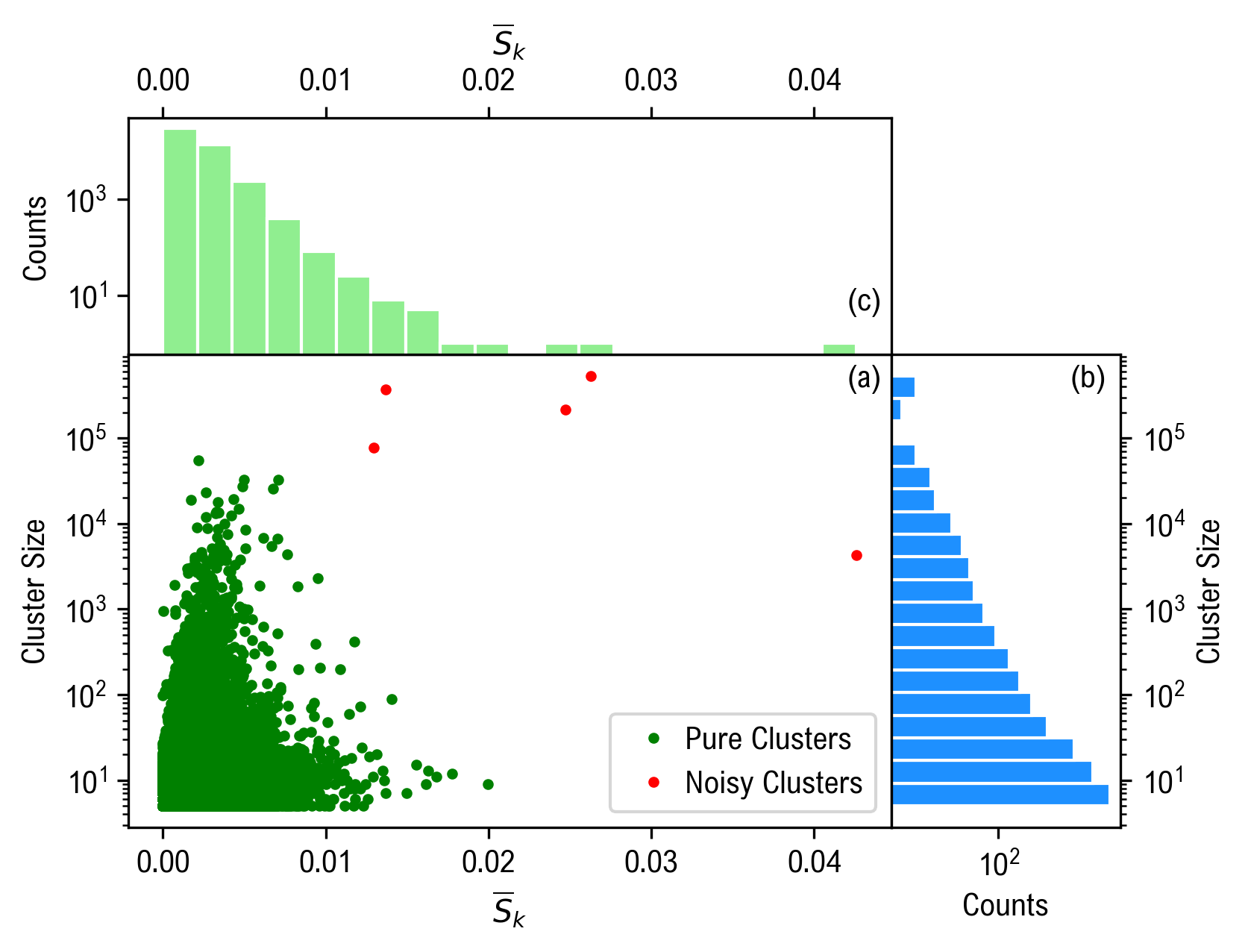}
    \caption{Panel (a): joint distribution of average-pairwise distance $\overline S$ and cluster size. Red dots represent "noisy clusters" identified by HDBSCAN and outliers with respect to the cluster size and $\overline {S}_k$  distributions, shown respectively in panels (b) and (c).}
    \label{fig:apwd}
\end{figure}

In our case, we employed the latter quantity to measure the purity of the clusters identified by HDBSCAN, together with the cluster size. In Figure~\ref{fig:apwd}(a) for each cluster is reported $\overline S_k$ and the cluster size. The red dots are the clusters that are not considered for the following analysis. Four of them correspond to the "noisy clusters" retrieved by HDBSCAN. All of them result in outliers with respect to the joint size and $\overline S_k$ distribution, whose marginals distributions are reported in panels (b) and (c) respectively.

\begin{figure}
     \centering
    \includegraphics[]{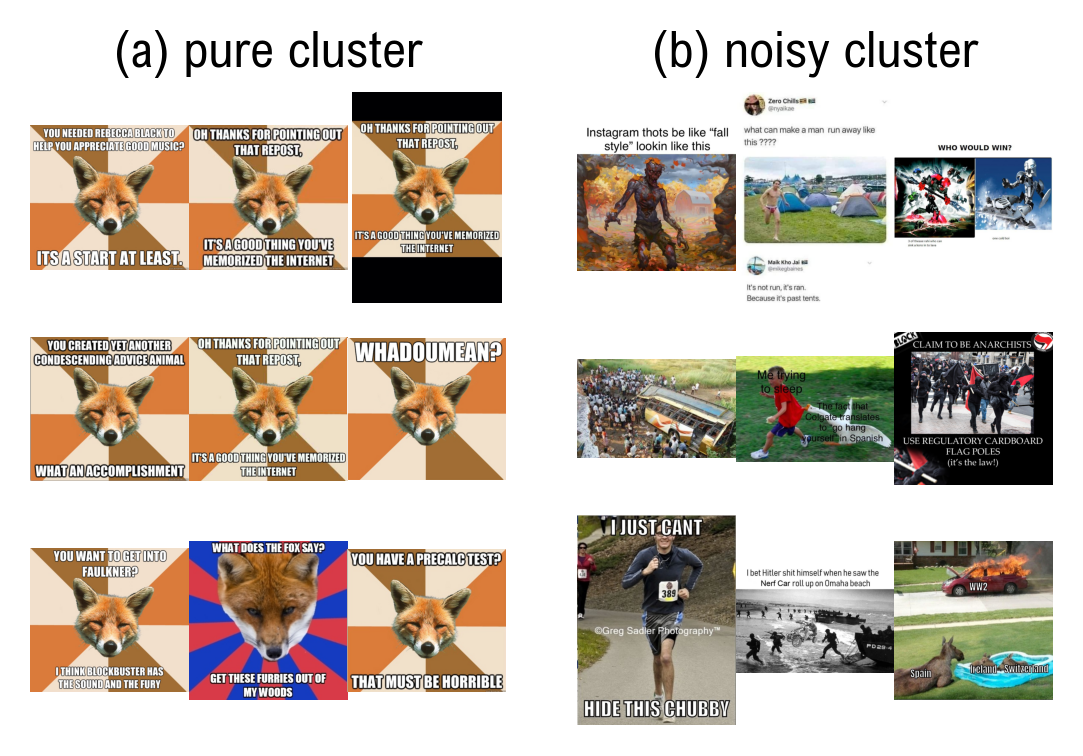}
    \caption{Examples of memes clusters. In a pure cluster, shown in panel (a), memes are different instances of the same template; insead, in a noisy cluster, shown in panel (b), it is not possible to detect a common template for all instances.}
    \label{fig:clust_examples}
\end{figure}

In Figure~\ref{fig:clust_examples} some memes from a pure cluster and a noisy one are reported. 

\subsection*{Entropy and complexity}
Permutation entropy $H$ and statistical complexity $C$ are two quantities that can be used to synthesize general properties of images, based on the value and relative disposition of their pixels. In the following, we give a minimal description of both quantities, and we refer the reader to original articles for more formal details \cite{Bandt2002,LpezRuiz1995,Ribeiro2012,Zunino2016,Rosso2007}. Permutational entropy measures the degree of disorder in the pixel arrangement. High values indicate high pixel randomness, while low values correspond to more regular patterns. Statistical complexity instead measures the amount of  “structural” complexity. Non-trivial spatial patterns give rise to positive values, while extremely ordered or disordered patterns correspond to low values. 

To compute $H$ and $C$ all colored images were converted to grayscale. Thus, each image consists of two-dimensional matrix. Next, following \cite{Sigaki2018}, for all the  $2\times 2$ submatrices comprised in the image the relative ordering of the pixel is computed. For a collection of four elements, a number of $4!=24$ possible permutations can be obtained. By counting the relative occurrence of each permutation, a probability mass function $P = \{p_1\dots p_N\}$ with 
\[
\sum_{i=1}^N p_i=1\,,
\]
can be built. Shannon's entropy is then computed over this probability distribution to obtain the permutation entropy
\begin{equation}
H(P) =\frac{S(P)}{\log(N)} =  \frac 1{\log(N)}\sum_{i=1}^N p_i\log\left(\frac 1{p_i}\right)\,.
\label{eqn:entropy}
\end{equation}
Given the probability mass function $P$, its discrepancy with respect to a uniform distribution $U = \{u_1\dots u_N\}$ with 
\[
\sum_{i=1}^N u_i=1\,,
\]
is obtained by computing the Jensen-Shannon divergence
\[
D(P,U)= S\left(\frac {P+U}2\right)-\frac{S(P)}2-\frac {S(U)}2\,.
\]
Combining this quantity with $H(P)$, the statistical complexity can be computed as
\begin{equation}
    C(P)=\frac{D(P,U)\, H(P)}{D^*}\,,
    \label{eqn:stat_compl}
\end{equation}
with the normalizing factor
\[
D^* = \max_P D(P,U) = -\frac 12\left[\frac{N+1}N\log(N+1)+\log(N)-2\log(2N) \right]\,.
\]

The evolution of visual patterns can be studied as a trajectory in the above defined entropy-complexity plane \cite{LpezRuiz1995, Ribeiro2012}, following the same approach used for paintings \cite{Sigaki2018}.

\section*{Results}
\label{sec:results}
Our study starts from Dawkins' hypothesis of the meme as the basic unit of cultural evolution, in connection with the post-memetics analyses of memes as cultural signs. We studied the evolution of Internet visual memes, i.e. images with (typically) overlapped text strings over a time span of 10 years. The dataset comprises around 2 million images, that were grouped together exploiting an unsupervised Machine Learning routine (see Methods). Such an extended dataset enables us to investigate some properties of this particular cultural meme. The clusters retrieved by our procedure correspond to the \textit{templates} of the various memes. In the following, we refer to "meme" as for the template, while each image belonging to a given template is an "instance" of the meme.

\begin{figure}
    \centering
    \includegraphics{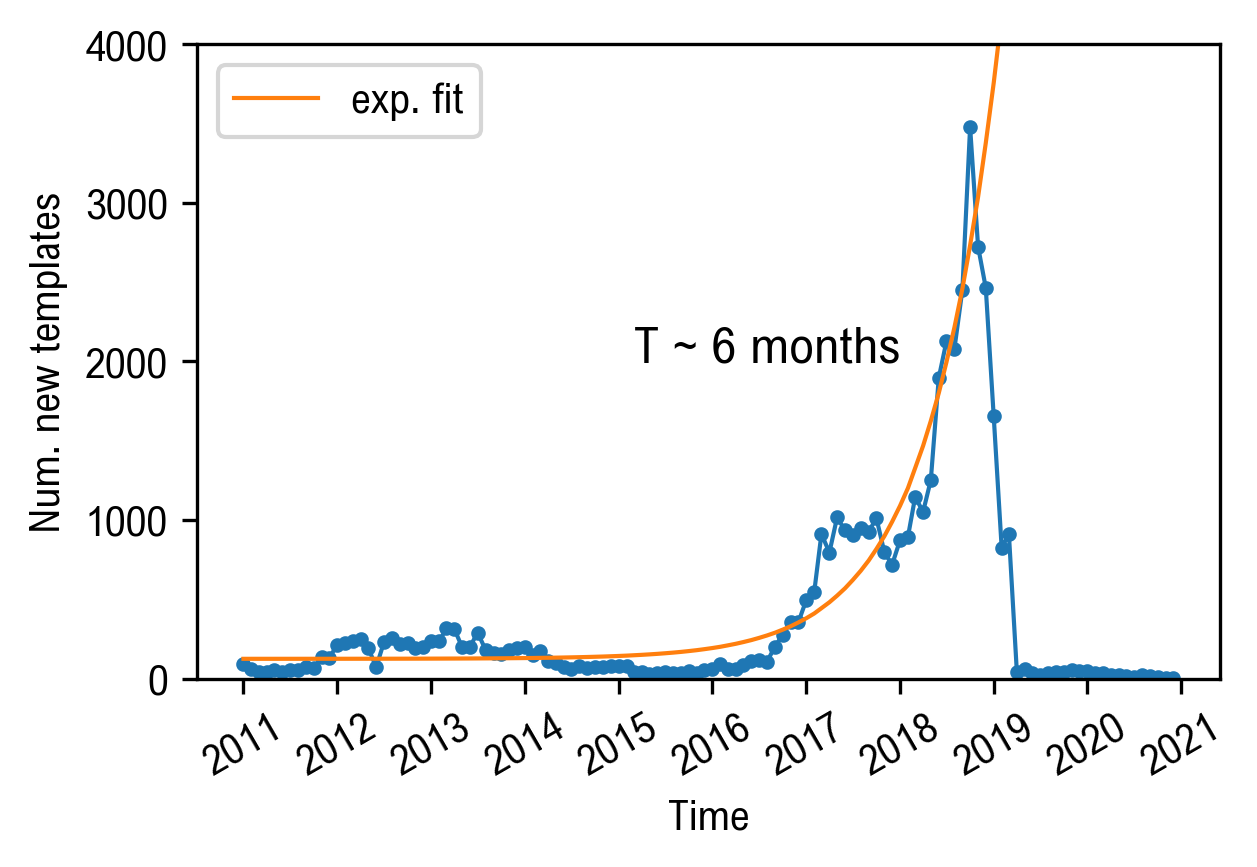}
    \caption{Evolution rate of internet memes. The number of new templates per each month is reported as a function of time. The growth rate is estimated through an exponential fit, with a doubling time of $T\sim 6$ months.}
    \label{fig:evolution_rate}
\end{figure}

To quantify the growth of memes adoption we computed their evolution rate. For each cluster, we store the creation time of its first instance. Next, per each week of sampling, we compute the number of new templates. The result is reported in Figure~\ref{fig:evolution_rate}, in which the growth rate is estimated by an exponential fit, giving a doubling time for the number of templates $T\sim 6$ months. The sudden drop in the plot is a finite size effect of the dataset. Namely, for subreddits \texttt{r/memes} and \texttt{r/dankmemes} it has not been possible to download the data over 2019, due to the exponential trend in the number of memes (see Figure~\ref{fig:dataset}). The fit was performed considering the data until January 2019.

\begin{figure}
    \centering
    \includegraphics{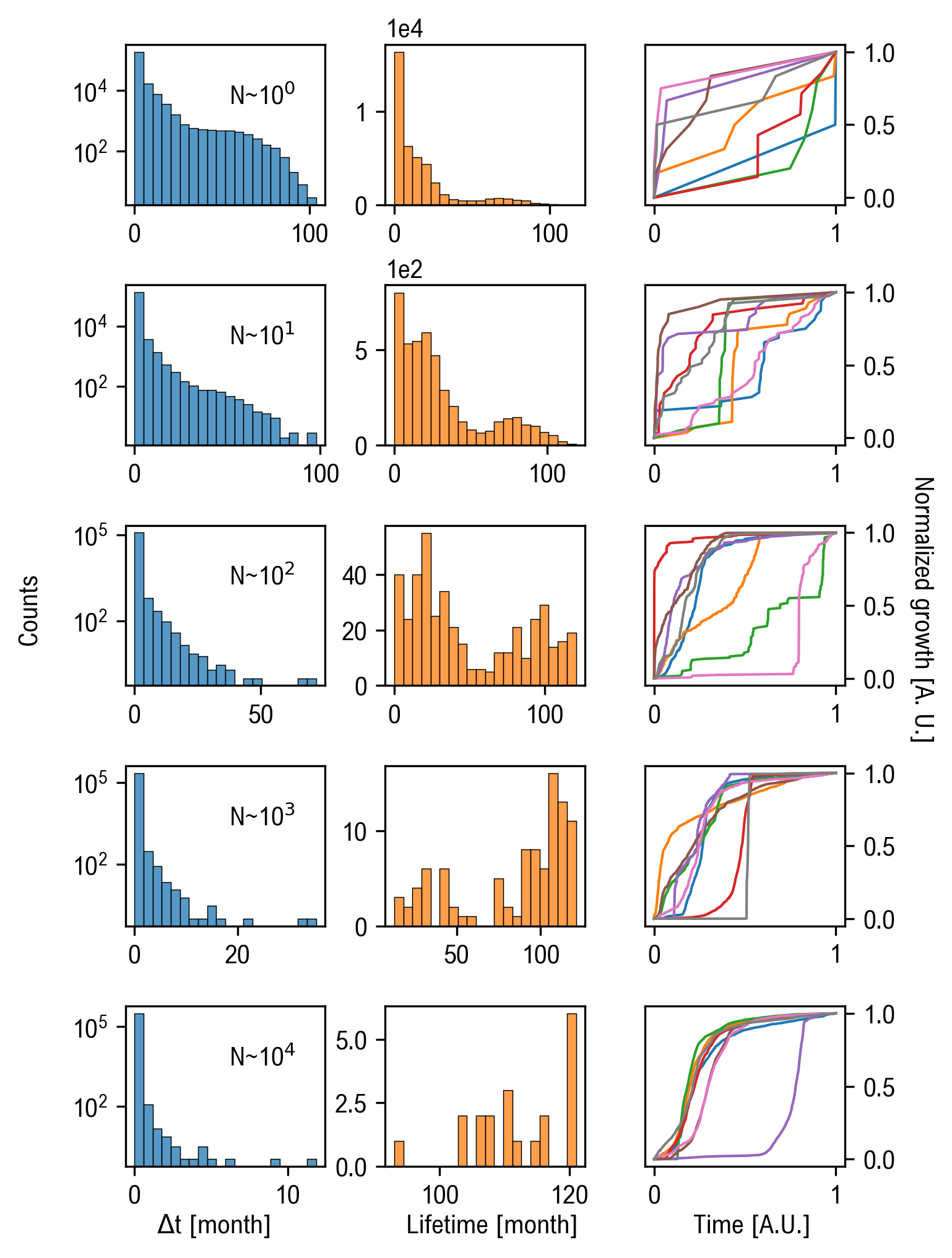}
    \caption{Mutation rate of memes. Left column: distribution of instances' inter-arrival times ($\Delta t$); central column: lifetime distribution of memes; right column: exemplary growth curve of meme adoption. Each row correspond to a typical size of meme cluster.}
    \label{fig:mutation_rate}
\end{figure}

Another relevant quantity in terms of cultural evolution is the \textit{mutation rate}. The mutation rate can be approximated by looking at the instances of each meme. For each template we computed the distribution of the differences ($\Delta t$) between the creation times of an instance and the following one. This distribution reveals the nature of growth of each cluster: a distribution skewed towards low values of $\Delta t$ corresponds to a very fast and bursty growth dynamic. Conversely, larger values of $\Delta t$ may reveal a more persistent template, whose instances occur more spaced in time. In Figure~\ref{fig:mutation_rate} the distribution of the "inter-instance" times (blue histograms, left column) is reported together with the clusters lifetime distribution (orange histograms, central column) , i.e. the time interval between the first and the last instance of a given meme. These distributions are computed for different typical cluster sizes ($N$).

Overall, lifetime results positively correlated with cluster size. Small clusters show a heterogeneous distribution of instances' inter-arrival times, comprising both small-bursty clusters and small-slowly paced ones. The lifetime is peaked towards low values. This behavior is also shown by very recent clusters whose actual size cannot be estimated within our dataset, as their evolution is ongoing.
As the cluster size grows, we observe a shift of the inter-times distribution towards low values, unveiling faster dynamics, while the lifetime distribution is concentrated around larger values. By looking the corresponding growth curves we observe an ensemble of trajectories that tend to display a fast initial build up of popularity, followed by a slower diffusion that determines the longer lifetime values. Conversely there are also examples of memes that takes more time to reach a wide popularity. This aspect may be due to non-trivial popularity dynamics, calling for further research.

\begin{figure}
    \centering
    \includegraphics{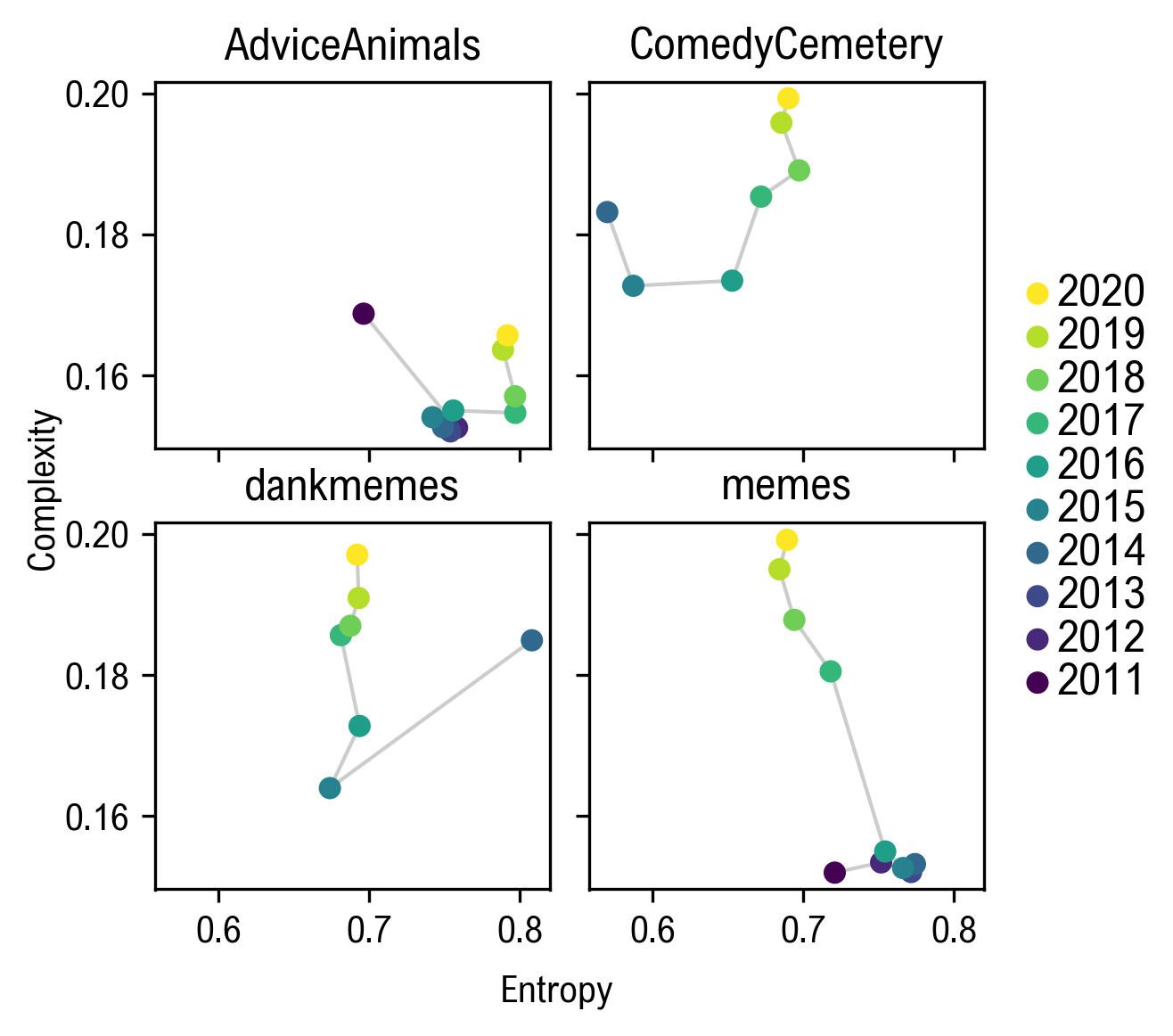}
    \caption{Trajectories in entropy-complexity plane for the four Reddit communities. All, except \texttt{r/AdviceAnimals} present an evolution towards higher values of complexity that resembles that of painting arts (see \cite{Sigaki2018}). Each dot is the average value of entropy and complexity for each year.}
    \label{fig:entropy_complexity}
\end{figure}

Following \cite{Sigaki2018}, we investigated the evolution of memes in the entropy-complexity plane. For each meme instance we computed the values of $H(P)$ and $C(P)$ and then averaged the obtained values by year. The results are reported, for each subreddit, in Figure~\ref{fig:entropy_complexity}. We observe that each community moves towards higher complexity values, except for \texttt{r/AdviceAnimals}, whose posting rules limit the natural evolution of produced memes; this effect could be also linked to the overall decrease in memes production observed for this community. Interestingly, also paintings followed a similar trajectory in the entropy-complexity plane: quite localized along the entropy axis, but shifting towards higher complexity in time (see. Figure 1 of ref. \cite{Sigaki2018}). 

The tendency of memes to evolve towards more complex structures can be explained considering this object as part of the emerging internet meta-language. In fact, memes are used to quickly vehicle context-specific content, which in turn evolves towards more and more specific templates. This may lead to a segregation effect, with a specific dialect depending on the community in which a meme is shared. In fact a meme created for a specific community, e.g. gaming community, does not have to be universally comprehensible across the web. This aspect leads to the use of more complex and specific patterns.

\section*{Discussion}
The Internet provides an environment in which information quickly spreads and adapts to comply with users' cognitive abilities.
A foundational question about memes is whether they present culturally and temporally transcendent characteristics in their organizing principles and how they evolve. 
Such a significant increase and spread of visual memes can be read under the light of post-memetics theories. Visual memes are favored by the rapid, fluid, continuously changing internet environment because of their simplicity, ease of handling and broad applicability in terms of subjects and situations. 
We find support for the hypothesis that memes are part of an emerging form of internet metalanguage: on one side, we observe an exponential growth with a doubling time around 6 months; on the other side, the complexity of memes contents increases, allowing to timely represent social trends and attitudes.
Our analysis shows that memes are relational entities functioning as flexible elements of a metalanguage that de-codifies and re-codifies the cultural system. They appear as fundamental components of an organic process that affects and conditions the digital environment and produces evolving forms of visual and textual complexity. 

\bibliography{scibib}

\bibliographystyle{Science}

\end{document}